\documentclass[12pt]{article}
\usepackage{graphics}
\usepackage{graphicx}

\begin{document}

\title{\Large \bf Precise measurement of  Br($K^+ \rightarrow \pi^+ \nu \bar{\nu}$) at CERN SPS }
\author{
\large Venelin Kozhuharov \footnote{On behalf of the P326 collaboration} 
 \\
  \bigskip \\
{\it University of Sofia 'St. Kliment Ohridski', Sofia, Bulgaria} \\
{\it JINR Dubna, Russia }
}

\maketitle

{
\large

\begin{center}
{\bf Abstract}\\
\medskip
\end{center}

An experimental proposal to measure  Br($K^+ \rightarrow \pi^+ \nu \bar{\nu}$) has been submitted at the CERN SPSC. The goal is to collect $\approx$80 $K^+\rightarrow\pi^+ \nu \bar{\nu}$ events 
with signal to background ratio of 10:1 in two years of data taking. This will allow determination of the CKM parameter $V_{td}$ with $\approx$10\% accuracy.

\section{Introduction} \label{intro}

Among the many kaon decays, the rare decays $K \rightarrow \pi \nu \bar{\nu}$ are extremely attractive as their branching ratios are theoretically very clean.
The reason is that the hadronic matrix element can be determined experimentally, using isospin symmetry of the strong interactions, from the leading semileptonic decay $K^+ \rightarrow \pi^0 e^+ \nu$ \cite{ISOSPIN_RELATION}. At the quark level $K \rightarrow \pi \nu \bar{\nu}$ decays arise from the $s\rightarrow d\nu\bar{\nu}$ transitions via $Z^0$ penguin and $W$ box diagrams. 

The branching ratio for $K^+ \rightarrow \pi^+ \nu \bar{\nu}$ can be written as \cite{BUCHALLA_BURAS_99}

\begin{equation}
Br(K^+\rightarrow\pi^+\nu\bar{\nu}) = k_+\left[\left(\frac{Im\lambda_t}{\lambda^5}X(x_t)\right)^2 + \left(\frac{Re\lambda_t}{\lambda^5}X(x_t) + \frac{Re\lambda_c}{\lambda^5}P_c\right)^2\right]
\label{branching}
\end{equation}
\begin{equation}
k_+=r_{K^+}\frac{3\alpha Br(K^+ \to \pi^0 e^+ \nu)}{2\pi^2 sin^4\theta_W}\lambda^8=(5.04\pm0.17)\times 10^{-11}\left(\frac{\lambda}{0.2248}\right)^8
\end{equation}
where $\lambda=V_{us}$, $\lambda_c=V_{cs}^*V_{cd}$, $\lambda_t=V_{ts}^*V_{td}$, $x_t=m_t^2/m_W^2$ and the $r_{K^+}=0.901$ is the isospin breaking correction. The coefficients $X$ and $P_c$ are being computed numerically and the most recent Next-to-Next-to-Leading Order $\chi PT$ calculation predicts $Br(K^+\rightarrow\pi^+\nu\bar{\nu})= (8.0 \pm 1.1)\times 10^{-11}$ \cite{BURAS_NNLO} where the error is almost entirely due to uncertainties in $m_c$ and the CKM matrix elements.
The present three events of this decay observed by E787 and E949 collaborations \cite{BNL_BR} lead to 
\begin{equation}
Br(K^+ \rightarrow \pi^+ \nu \bar{\nu})=(14.7^{+13.0}_{-8.3})\times 10^{-11}
\end{equation}

Investigation of  both $K^+ \rightarrow \pi^+ \nu \bar{\nu}$ and $K^0 \rightarrow \pi^0 \nu \bar{\nu}$ provides an alternative way for determination of the apex of the unitarity triangle. The comparison between this apex and the one obtained from the $B$-mesons gives possibly the cleanest test of the Standard Model.

\section{Measurement strategy}

In $K^+ \rightarrow \pi^+ \nu \bar{\nu}$ decay there is only one observable particle in the final state, $\pi^+$, therefore we need to use two spectrometers to measure the kaon momentum as well as the pion momentum.
\begin{figure}
\begin{center}
\includegraphics[width=7truecm]{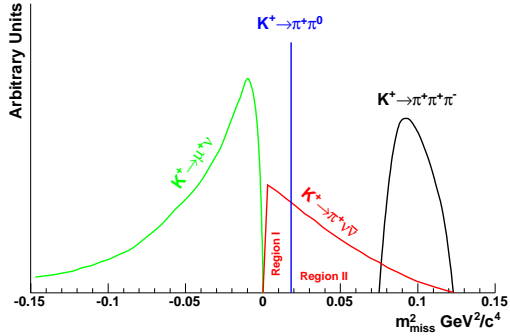}
\caption{Distribution of the missing mass squared for $K^+ \rightarrow \pi^+ \nu \bar{\nu}$, $K^+ \to \mu^+ \nu$, $K^+\to \pi^+ \pi^0$ and $K^+\to \pi^+\pi^+\pi^-$ decays} 
\label{missing_mass}
\end{center}
\end{figure}
Under the hypothesis that the observed charged particle in the final state is pion and that the decaying particle is kaon the missing mass squared is given by

\begin{equation}
m_{miss}^2 \simeq m_K^2\left(1-\frac{|P_{\pi}^2|}{|P_{K}^2|}\right)
+ m_{\pi}^2\left(1-\frac{|P_{K}^2|}{|P_{\pi}^2|}\right) - |P_{K}||P_{\pi}|\theta_{\pi K}^2
\label{M_MISS}
\end{equation}

On Figure \ref{missing_mass} the missing mass squared distribution for the three most frequent $K^+$ decays together with $K^+ \rightarrow \pi^+ \nu \bar{\nu}$ is shown. Since the peak of $K^+\rightarrow \pi^+\pi^0$ decays lies in the signal region we separate the signal into two different regions
\begin{itemize}
\item Region I:  $0 GeV^2/c^4 < m_{miss}^2 < 0.01 GeV^2/c^4 $ 
\item Region II: $0.026GeV^2/c^4 < m_{miss}^2 < 0.068 GeV^2/c^4 $
\end{itemize}
with borders defined by the resolution on the $m_{miss}^2$. 

In order to suppress the background in the defined kinematic regions particle identification and a system of veto detectors will be used. The detector should be made hermetic for photons originating from $\pi^0$ in the decay region. A cut on the maximum $\pi^+$ momentum of 35 GeV will be made to assure that the energy deposited in the photon veto system will be at least 35GeV. The overall photon veto system has to provide an inefficiency  less than $10^{-8}$ for a $\pi^0$ coming from $K^+\rightarrow \pi^+\pi^0$ decay. Decays with muons in the final state (like $K^+\rightarrow \mu^+ \nu$, $K^+\rightarrow \pi^+\pi^-\mu^+ \nu$ ) will be suppressed using a muon veto system for which the total inefficiency should be less less than  $5 \times 10^{-6}$.

\section{Experimental setup} \label{setup}
The overall beam and detector layout is designed to meet all the requirements defined and is shown on Figure \ref{detector}. Kaon decays in flight technique is considered. All mentioned angles are with respect to the Z-axis of the experiment (also used as a beam axis shown with dashed line on the Figure \ref{detector}) if not specified explicitly.

\begin{figure}[t]
\begin{center}
\includegraphics[width=14truecm]{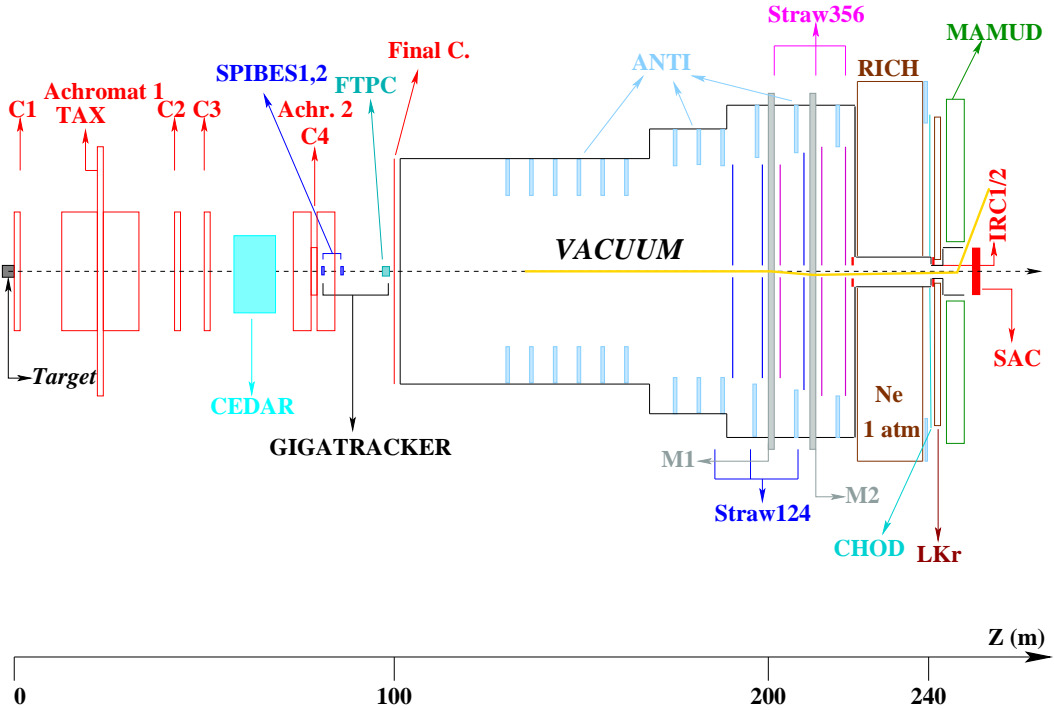}
\caption{P326 detector layout}
\label{detector}
\end{center}
\end{figure}

\subsection{Beam}
The  beam of positive charged particles is produced by 400 GeV primary proton beam with an intensity of $3\times 10^{12}$ protons per 5 s pulse hitting a beryllium target of 2 mm diameter and 40 cm length at a zero incidence angle. The particles with central momentum of 75 GeV and a momentum bite of 1\% are selected by an achromat and transported through a differential Cherenkov counter. This counter will be able to tag only $K^+$ which represent $\approx$6\% of the beam content. A second achromat which serves also as a beam spectrometer is placed in the next 30 m before the final collimator. 

 The selected particles enter the decay region which is housed in a $\approx$120 m long evacuated tank closed by a thin aluminium window. The center of this window is traversed by a beam tube allowing the beam to be transported in vacuum through the detectors downstream of the tank.

Within the detectors following the decay region the beam is deflected by two large-aperture dipole magnets of a double spectrometer. The beam is centered again on the axis at the place where a combined magnetized-iron hadron calorimeter and muon detector (MAMUD) is located. The MAMUD is designed to produce a +1500 MeV/c $p_T$-kick on the beam
thus clearing a 100 mm radius photon veto calorimeter located 8 m further downstream.

\subsection{Detector components}

\subsubsection{Tagging subsystem}
\begin{itemize}
\item CEDAR

Positive $K^+$ tagging helps to reduce the background from the beam pions interacting in the residual gas of the vacuum tank. It is planned to use an upgraded version of one of the existing  differential Cherenkov counters at CERN SPS - CEDAR West. This instrument will function well for our purposes if filled with hydrogen thus minimizing the scattering of the beam in the gas.

\item CHOD 

The charged hodoscope is intended to be used in the event trigger and therefore must possess a very good efficiency for detecting charged particles. The hodoscope signal will be used offline to guarantee that the charged track is properly associated with the incoming particle. For this reason it should provide an intrinsic time resolution better than 100 ps. We will use multi-gap glass RPCs \cite{PROP_46} which have shown time resolution  down to 50 ps with rates up to 1 $kHz/cm^2$ and efficiency above 99.9\% during tests \cite{PROP_47}. 

\end{itemize}

\subsubsection{Tracking subsystem}
\begin{itemize}

\item GIGATRACKER

 The beam spectrometer should perform measurement of the beam parameters at a rate of 1 GHz. The beam size will be $3.2\times 4.4$ cm which leads to an average particle rate of $60 MHz/cm^2$. 
A hybrid detector is considered consisting of two stations of thin fast silicon micro-pixel detectors (SPIBES) for redundant momentum measurement of the beam and a micromegas-based Time Projection Chamber (FTPC) to measure the direction of the incoming particles with minimum amount of material.
The SPIBES will provide also time resolution of the order of 100 ps which will simplify the pattern recognition in the FTPC. The ultimate requirements for SPIBES can be achieved with $300 \mu m$ thick detector coupled to $100 \mu m$ readout chip for a total of  $\approx 0.4\% X_0$ material per station. The FTPC will be an improved version of the 
gaseous beam detector developed for NA48/2 \cite{NA48}
equipped with FADC read-out. 
The precision of the beam momentum measurement is expected to be at the level of $0.3\%$.

\item STRAW SPECTROMETER

In order to provide redundant measurements of the outgoing pion momentum and angle a double magnetic spectrometer is considered. It will consist of two dipole magnets with two chambers between them, two chambers before the first magnet and two chambers after the second magnet. Chambers of straw tubes will be used because of their ability to operate in vacuum thus minimizing the multiple scattering. The amount of material for each four-view chamber crossed by a track corresponds to a 0.4\% of a radiation length
. The momentum resolution is expected to be better than $1\%$ for 30 GeV/c pion. 

\end{itemize}

Such a tracking system will provide a measurement of the relative angle between the kaon and the pion with precision of $50-60 \mu rad$.

\subsubsection{Photon veto subsystem}
In order to suppress the background coming from decays with $\pi^0 $ in the final state a system of photon veto counters is envisaged.
\begin{itemize}
\item ANTI 

ANTIs are a set of ring shaped anti-counters surrounding the decay region. The tank will be equipped with 13 counters, each composed of 80 layers of 5 mm scintillator and 1 mm lead (16 $X_0$) with wave-length shifting (WLS) fibers readout. They will provide full coverage for photons originating from the decay volume with angles up to 50 mrad.
 
\item LKR

The existing NA48 Liquid Krypton calorimeter will be exploited as a photon veto detecting photons from the decay volume with angles between 1.0 mrad and 15.0 mrad. The excellent performance of the calorimeter \cite{PROP_35} will also help to reject the background from $K^+ \rightarrow \pi^0 e^+ \nu$ decays. An upgrade of the read-out electronics is considered in order to achieve single photon inefficiency of $10^{-5}$ for photons with energy above 5 GeV.

\item IRC1/2, SAC

The two sets of small angle vetoes covering the regions around and in the beam will be 17 $X_0$ deep in order to keep the photon punch through probability bellow $10^{-7}$. An appropriate solution is based on shashlyk technique \cite{PROP_37} of 100 layers of 1 mm lead and 2 mm scintillator with longitudinal WLS fibers.

\end{itemize}

\subsubsection{Muon veto subsystem}

\begin{itemize}

\item RICH

A gas Ring Imaging Cherenkov counter will provide muon/pion separation more than three sigma for momentum up to 35 GeV/c. The design chosen is similar to the KPLUS \cite{PROP_45} one adapted to our layout. The RICH can be also used as a velocity spectrometer.

\item MAMUD

The detector is designed as a an iron scintillator sandwich and  consists of 150 iron plates, 2cm thick and $2.8m\times2.6 m$ size, interspersed with extruded scintillators. It will be equipped with two coils which will produce about 1 T magnetic field on the axis. Such a detector is capable of identifying muons with inefficiency of the order of $10^{-5}$. It also serves the purpose of deflecting the charged beam away from the photon detector (SAC) at the end of the hall. 

\end{itemize}

\section{Conclusion}

With the presented setup and analysis technique the acceptance for the signal is about $10\%$ with signal to background ratio of 10:1. Assuming two years of data taking with total of $4.8\times 10^{12}$ kaon decays per year in the fiducial volume and the Standard Model prediction for $K^+ \rightarrow \pi^+ \nu \bar{\nu}$ branching ratio 80 signal events are expected to be observed. 

The collaboration has been established and a proposal has been submitted to SPSC\cite{P326_PROP}. The experiment is planned to take data in 2009 and 2010.

}


\begin{thebibliography}{99}


\bibitem{ISOSPIN_RELATION}
  W.~J.~Marciano and Z.~Parsa,
  Phys.\ Rev.\ D {\bf 53}, 1 (1996).


\bibitem{BUCHALLA_BURAS_99} 
  G.~Buchalla and A.~J.~Buras,
  Nucl.\ Phys.\ B {\bf 548} (1999) 309
  [arXiv:hep-ph/9901288].

\bibitem{BURAS_NNLO} 
  A.~J.~Buras, M.~Gorbahn, U.~Haisch and U.~Nierste,
  arXiv:hep-ph/0508165.

\bibitem{BNL_BR}
  S.~Adler {\it et al.}  [E787 Collaboration],
  Phys.\ Rev.\ Lett.\  {\bf 88} (2002) 041803
  [arXiv:hep-ex/0111091], 
  V.~V.~Anisimovsky {\it et al.}  [E949 Collaboration],
  Phys.\ Rev.\ Lett.\  {\bf 93} (2004) 031801
  [arXiv:hep-ex/0403036].



\bibitem{NA48}
  NA48 Collaboration, http://na48.web.cern.ch/NA48/


\bibitem{PROP_46}
  A.~Akindonov {\it et al.},
  Nucl.\ Instrum.\ Meth.\ A {\bf 456} (2000) 16.
  [arXiv:hep-ph/9901288].

\bibitem{PROP_47}
  A.~Akindonov {\it et al.},
  Nucl.\ Instrum.\ Meth.\ A {\bf 533} (2004) 74.
  [arXiv:hep-ph/9901288].

\bibitem{PROP_35}
  G.~Unal, 
  Performance of the NA48 Liquid Krypton Calorimeter, Proceedings of the $9^th$ International Conference on Calorimetry in High Energy Physics, 
  Annecy, Oct. 2000, p. 355.

\bibitem{PROP_37}
  G.~S.~Atoian {\it at al.},
  Nucl.\ Instrum.\ Meth.\ A {\bf 531} (2004) 467.
  [arXiv:physics/0310047]

\bibitem{PROP_45}
  P.~S.~Cooper [the CKM Collaboration], ``Redesign of the CKM RICH velocity spectrometers for use in a 1/4-GHz beam''. FERMILAB-CONF-05-015-CD 


\bibitem{P326_PROP}
  G.~Anelli {\it et al.},
CERN-SPSC-2005-013 

\end{thebibliography}
\end{document}